\documentclass[prb,epsf,twocolumn,showpacs,preprintnumbers]{revtex4}
\usepackage{graphics}
\usepackage{graphicx}
\usepackage{dcolumn} 
\usepackage{bm}
\usepackage{epsfig}
\usepackage{float}
\begin{document}
\title{Crystal water induced switching of magnetically active orbitals in CuCl$_2$}
\author{M.~Schmitt$^1$, O.~Janson$^1$, 
  M.~Schmidt$^{1}$, S.~Hoffmann$^1$, W.~Schnelle$^1$,
  S.-L.~Drechsler$^{2}$ and H.~Rosner$^1$}
\affiliation{$^1$Max-Planck-Institut f\"ur Chemische Physik fester
Stoffe, 01187 Dresden, Germany}
\affiliation{$^2$IFW Dresden, P.O.Box 270116, 01171 Dresden, Germany}
\begin{abstract}
The dehydration of CuCl$_2$$\cdot$2H$_2$O to CuCl$_2$ leads to a
dramatic change in magnetic behavior and ground state.
Combining density functional electronic structure and model
calculations with thermodynamical measurements we reveal the
microscopic origin of this unexpected incident -- a crystal water
driven switching of the magnetically active orbitals. This switching
results in a fundamental change of the coupling regime from a
three-dimensional antiferromagnet to a quasi one-dimensional behavior.
CuCl$_2$ can be well described as a frustrated $J_1$--$J_2$ Heisenberg
chain with ferromagnetic exchange $J_1$ and $J_2/J_1 \sim -1.5$ for
which a helical ground state is predicted.
\end{abstract}

\maketitle

\section{Introduction}
Low-dimensional spin 1/2 magnets are of wide interest in solid state
physics since they are ideal objects to study the interplay of
dimensionality, magnetic frustration and strong quantum fluctuations.
These compounds can be described often very successfully based on
their magnetically active structural building blocks and their
linking. Typical examples for such building blocks are Ti(III)O$_6$
octahedra, V(IV)O$_5$ square pyramids or Cu(II)O$_4$ plaquettes that
form various spin 1/2 networks like quasi one-dimensional (1D) chains
or ladders.  Nevertheless, for a reliable and accurate description of
such networks, precise model parameters are a precondition, especially
in the vicinity of quantum critical points. However, since for new
compounds these parameters are unknown, it is common to transfer the
known parameters from related, similar systems in a slightly
renormalized form according to changed distances and/or bond angles.

A typical example for the application of this strategy are compounds that 
contain crystal water in different amounts.
Although in some cases the topology of the magnetic network and the
related magnetic properties totally change upon dehydration like in the
case of CuSiO$_3$$\cdot$H$_2$O~\cite{gros02,heide55} and
CuSiO$_3$~\cite{rosner01,otto99}, for most compounds only moderate
structural changes with respect to the magnetic network are observed.
It is generally assumed that in this case crystal water leads mainly
to a modest change of the crystal field for the magnetic ion.  In turn,
small changes in the crystal field only, would directly suggest a
description within the same model with slightly revised
parameters.\cite{janson08} This leads to the common believe that
crystal water plays only a minor role regarding the magnetic
properties for compounds where crystal structure is basically preserved 
upon water intercalation.

Here, in contrast, we show that the hydration of CuCl$_2$ to
CuCl$_2$$\cdot$2H$_2$O fundamentally changes the magnetic properties,
although the topology of the covalent Cu-Cl network is seemingly
unchanged. Whereas CuCl$_2$$\cdot$2H$_2$O is a classical
three-dimensional (3D) antiferromagnet (AFM) with a N\'eel temperature
of 4.3\,K, we establish the dehydrated species as an example for a
quasi 1D chain compound. The results of susceptibility measurements,
density functional and model calculations can be consistently
understood from a reorientation of the magnetically active Cu orbital
driven by the hydration. Regarding the well known 3D magnetic nature
of CuCl$_2$$\cdot$2H$_2$O,\cite{marshall, handel} the quasi 1D
behavior of the water free compound CuCl$_2$ is rather surprising:
CuCl$_2$ is a $J_1$--$J_2$ Heisenberg chain model compound with FM
nearest neighbor (NN) exchange $J_1$ and AFM next-nearest neighbor
(NNN) exchange $J_2$. According to the estimated $J_2/J_1\sim -1.5$ we
predict a helical magnetic order below the observed transition at
24\,K. Earlier studies \cite{stout,deJongh} that tried to model
CuCl$_2$ as a spin 1/2 chain found considerable deviations from a 1D
behavior since in these investigations only NN AFM coupling was
considered. Alternatively a 1D model with AFM NN and NNN exchanges was
suggested, but this model yields far too large couplings.\cite{banks}

\begin{figure}[tbh]
\begin{center}\includegraphics[%
  clip,
  width=7.5cm,
  angle=0]{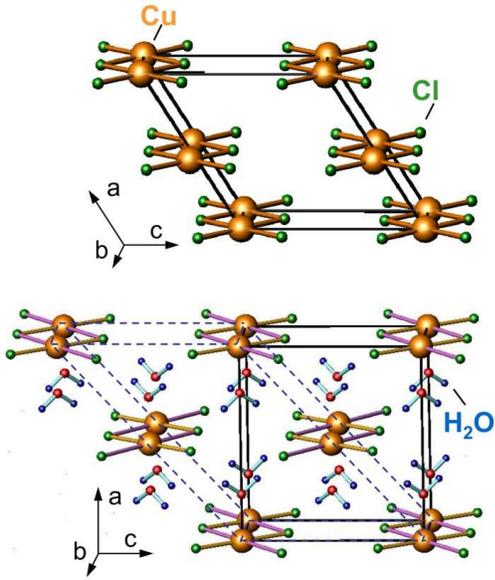}\end{center}
\caption{\label{bonds}(color online) 
Crystal structures of CuCl$_{2}$ (top) and CuCl$_2$$\cdot$2H$_2$O
(bottom). The shortest Cu--Cl bonds are highlighted (brown). The crystal
water is intercalated between the CuCl$_2$ layers. 
The dashed line in the  CuCl$_2$$\cdot$2H$_2$O structure marks the
unit cell of the water-free  CuCl$_2$.
}
\end{figure}

\section{Crystal Structure}
The crystal structure of CuCl$_2$ is presented in Fig.~\ref{bonds}. Cu
and Cl form a covalent network of edge shared CuCl$_4$ plaquettes
running along $b$ direction. Such fourfold planar Cu$^{2+}$
coordination suggests a strong analogy to the undoped cuprates.  The
Cu-Cl-Cu bond angle is 93.6$^\circ$ and thus very similar to that in
the CuO$_2$ chain cuprate family, where bond angles close to
90$^\circ$ result in FM NN exchange like in
Li$_2$CuO$_2$,\cite{wangbo,nitzsche} LiCu$_2$O$_2$\cite{gippius,
masuda, drechsler05} and Li$_2$ZrCuO$_4$
\footnote{These compounds are known as $J_1$--$J_2$ chain models with
  FM NN and AFM NNN exchange, where the different $J_2/J_1$ ratios
  lead to different ground states.}.\cite{drechsler}
As in the latter two compounds, these chains are arranged in layers
(Fig.~\ref{bonds}, top), suggesting a rather weak exchange
between the layers. Since even the arrangement of the chains within
the layers is very similar, a quasi 1D behavior might be expected
from a mere comparison with these cuprate crystal structures.

When CuCl$_2$ is exposed to moisture, H$_2$O enters the space between the chain
layers, finally forming the fully hydrated CuCl$_2$$\cdot$2H$_2$O 
(Fig.~\ref{bonds}, bottom). Although the crystal structure seems very similar
at a first glance, the crystal water
induces several changes: (i) The inter-layer distance increases, (ii)
the CuCl$_2$ chains shift with respect to each other and, (iii) the
Cu-Cl distances within the chains are modified. Whereas the structural
changes can be easily understood by packing and electrostatics
(negatively polarized O$^{2-}$ is situated close to the Cu$^{2+}$
ions, H$^+$ is attracted by Cl$^-$), the origin of the drastic change
of magnetic properties -- 1D versus 3D -- is far from
obvious. Unraveling the underlying microscopic physics is the aim of
our joint theoretical and experimental study.

\section{Methods}
Polycrystalline CuCl$_2$ was prepared by dehydration of
CuCl$_2$$\cdot$2H$_2$O (Alfa Aesar 99.999\%) 
under vacuum at
390\,K. Single crystals were grown by chemical transport
in a temperature gradient from 650\,K to 575\,K
with AlCl$_3$ (Alfa Aesar 99.999\%, ultra dry) 
as transport
agent. The chemical characterization of CuCl$_2$$\cdot$2H$_2$O
and of the CuCl$_2$ crystals was carried out by X-ray
powder diffraction, DSC/TG-methods and chemical analysis. The heat of
dehydration was determined by DSC.\footnote{Netzsch DSC 204
thermocouple, platinum crucible, heating rate
10\,K/min, dry argon atmosphere, 10\,mg sample.} 
Based on five independent measurements, the heat of
dehydration is $\Delta H^0_{\mathrm{dehyd.}} = (117\pm 2)$\,kJ/mol at 400\,K.

Magnetization was measured in a SQUID magnetometer ($1.8 - 300$\,K) in
magnetic fields up to 7\,T. Heat capacity was determined by a relaxation method
in the same temperature range up to $\mu_0H = 9$\,T.

Exact diagonalization (ED) of the $J_1$--$J_2$ Heisenberg Hamiltonian has
been performed on $N=16$ sites clusters using the ALPS
code.\cite{ALPS}  The low-temperature behavior of the magnetic
susceptibility has been simulated using the transfer-matrix
density-matrix renormalization-group (TMRG) method.\cite{tmrg}

For the electronic structure calculations the full-potential
local-orbital scheme FPLO (version: fplo7.00-28) within the local
(spin) density approximation (L(S)DA) was used.\cite{fplo1} In the
scalar relativistic calculations the exchange and correlation
potential of Perdew and Wang was chosen.\cite{PW92} To consider the
strong electron correlations for the Cu 3$d^9$ configuration, we use
the LSDA+$U$~\cite{fplo2} approximation varying $U_{d}$ in the
physically relevant range from 6 -- 8.5\,eV.  The LDA results were
mapped onto an effective tight-binding model (TB) and subsequently to
a Hubbard and a Heisenberg model.

\begin{figure}[t]
\begin{center}\includegraphics[%
  clip,
  width=8.5cm,
  angle=0]{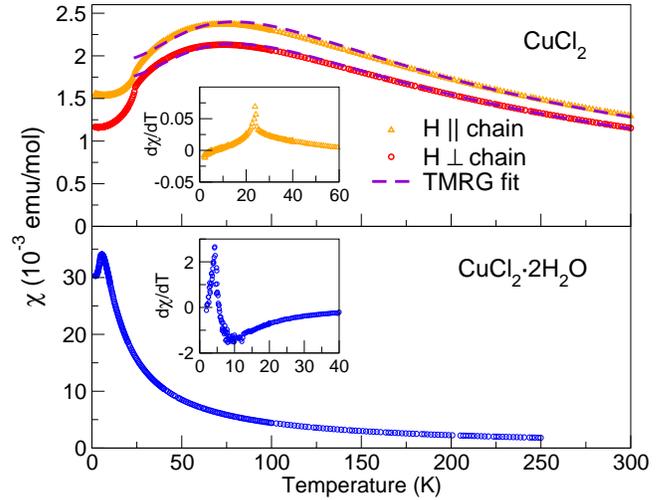}\end{center}
\caption{\label{sus}(color online)
Magnetic susceptibility of single crystalline CuCl$_{2}$ (top)
and CuCl$_2$$\cdot$2H$_2$O powder (bottom) as a function of
temperature ($\mu_0H=1$\,T). TMRG fits are given by the dashed lines. 
The insets show the derivative $d\chi/dT$, the ordering temperatures are
 indicated by sharp peaks.}
\end{figure}

\begin{figure}[t]
\begin{center}\includegraphics[%
  clip,
  width=8.5cm,
  angle=0]{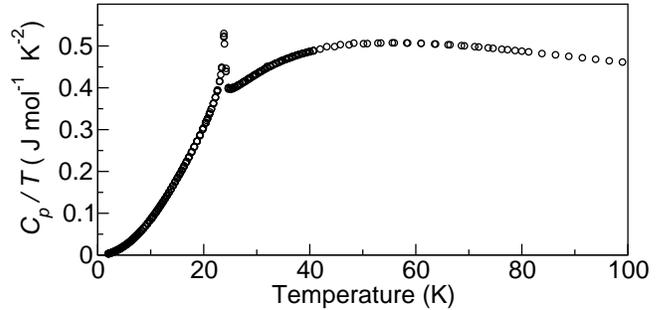}\end{center}
\caption{\label{cp}
Temperature dependence of the specific heat $C_p/T$ of single crystalline CuCl$_{2}$.}
\end{figure}
 
\section{Results and Discussion}
\subsection{Thermodynamic measurements}
The susceptibility data for both compounds are shown in
Fig.~\ref{sus}. CuCl$_2$ exhibits a broad maximum at $T_{\mathrm{max}}\approx
75$\,K as a fingerprint of quasi 1D behavior
\footnote{The absence of an impurity related Curie tail at low
  temperatures indicates the high quality of the sample.}.
An AFM Curie-Weiss temperature $\Theta_{\mathrm{CW}} = +107$\,K has been
extracted from the high temperature region. A sharp kink at $T_N =
24$\,K (see Fig.~\ref{sus} and inset $d\chi/dT$) followed by a rapid
drop of $\chi$ indicates a magnetic phase transition as earlier
suggested.\cite{stout} 

The measured zero-field specific heat as a function of temperature for CuCl$_2$ is shown 
in Fig.~\ref{cp}. Our data agree well with earlier studies.\cite{stout, banks}
The specific heat curve shows a pronounced lambda shape anomaly at 
$T_N = 24$\,K due to the onset of long range AFM ordering. The ordering temperature
from $C_p$ is in perfect agreement with $T_N$ evaluated from susceptibility. 
Well below $T_N$ the total specific heat is described by $C_p = \beta T^3$ with 
$\beta =$~0.8514(3)\,mJ mol$^{-1}$ K$^{-1}$ (fit for $T <$~10\,K), indicating that both 
phononic and magnetic contributions to $C_p$($T$) are $\propto T^3$. The data
in a field $\mu_0H = 9$\,T show no visible differences to the zero-field data.

In contrast to the quasi 1D susceptibility of CuCl$_2$, the hydrated
system shows an increasing $\chi$ down to low temperatures right above
the AFM phase transition at 4.3\,K (see Fig.~\ref{sus} and inset
$d\chi/dT$) in perfect agreement with earlier
measurements.\cite{marshall, handel} A Curie-Weiss temperature
$\Theta_{\mathrm{CW}} = +5.3$\,K indicating weak AFM interactions has been evaluated.

\subsection{Band structure calculations}
The essentially different character of the susceptibility of both compounds
points to a changed coupling
regime rather than to a mere re-scaling according to the modified
atomic distances. To construct an appropriate microscopic model based
on the relevant interactions we perform 
{\it ab-initio} electronic structure calculations, as successfully
demonstrated earlier for the closely related CuO$_2$ chain compound
family \cite{wangbo,nitzsche,gippius,drechsler}.

\begin{figure}[t]
\begin{center}\includegraphics[%
  clip,
  width=8.0cm,
  angle=0]{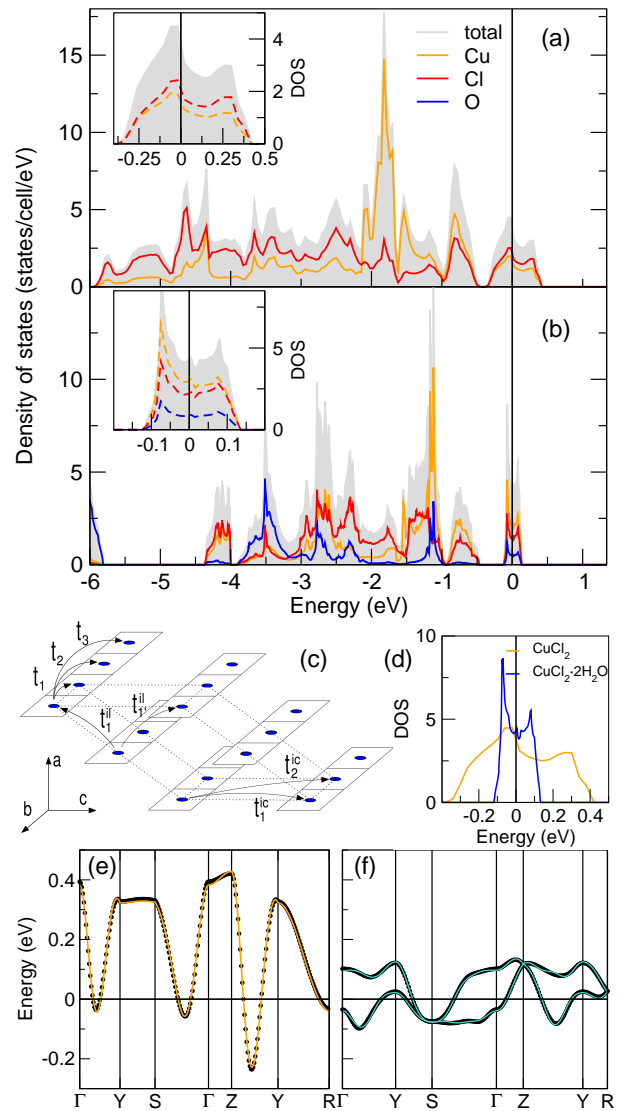}\end{center}
\caption{\label{dos}(color online) 
Total and partial 
density of states and band structure of the anti-bonding band of
CuCl$_{2}$ (a,e) and CuCl$_2$$\cdot$2H$_2$O (b,f). The insets show the
orbital character of the anti-bonding Cu-Cl (a) and Cu-Cl-O (b)
states, respectively. The bandwidths are compared in (d). 
The fits (full lines (e,f)) of the TB model (c) are
superimposed to the LDA band structures (circles (e,f)). The two bands
in (f) originate from the two Cu atoms per cell in
CuCl$_2$$\cdot$2H$_2$O.}
\end{figure}

Total and partial densities of states (DOS) for both
compounds are pictured in Fig.~\ref{dos}. On a coarse energy scale
both systems are similar, the contribution of the additional H
states to the valence region is negligible. Both compounds show half-filled, 
well separated anti-bonding bands at the Fermi level. This
metallic behavior is in contrast to the experiment and is a well known
shortcoming of the LDA due to the underestimation of the strong
Coulomb repulsion for the Cu$^{2+}$ 3$d^9$ configuration.  The
observed insulating ground state is obtained (i) within the LDA+$U$
approximation or (ii) by a model approach mapping the relevant low
lying LDA states onto an effective TB model, and, including the
correlations, subsequently onto a Hubbard and a Heisenberg model.

\begin{figure}[t]
\begin{center}\includegraphics[%
  clip,
  width=7.5cm,
  angle=0]{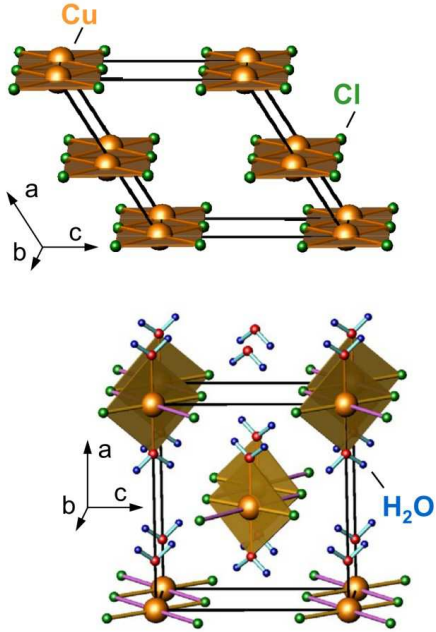}\end{center}
\caption{\label{orbitals}(color online) Sketch of the magnetically
  active $pd\sigma$-plaquette for CuCl$_{2}$ (top) and
  CuCl$_2$$\cdot$2H$_2$O (bottom). Whereas these orbitals form quasi 1D
  edge shared chains in CuCl$_2$, the plaquettes in
  CuCl$_2$$\cdot$2H$_2$O are isolated resulting in weak, but 3D
  interactions.}
\end{figure}

A closer inspection of the DOS and the related band structure reveals
two important differences: (i) Whereas the width of the anti-bonding
band in CuCl$_2$ is 0.8\,eV -- rather typical for 1D edge-shared
CuO$_2$ chains \footnote{For comparison, CuGeO$_3$ shows a bandwidth of
  0.95\,eV for the anti-bonding band, CuSiO$_3$ about 0.65\,eV and
  LiCuVO$_4$ about 0.7\,eV.} -- the bandwidth in CuCl$_2$$\cdot$2H$_2$O
is reduced by more than a factor of three to about 0.25\,eV
(Fig.~\ref{dos}d). (ii) The magnetically active
anti-bonding band in CuCl$_2$ is formed exclusively by Cu-Cl $dp\sigma$ states
(inset Fig.~\ref{dos}a) corresponding to the bonds pictured in
Fig.~\ref{bonds}.  In contrast, for CuCl$_2$$\cdot$2H$_2$O the O 2$p$
orbitals that are directed towards the Cu contribute significantly to
this band (inset Fig.~\ref{dos}b). This leads to the formation of a
new $dp\sigma$ orbital perpendicular to the original ones. The related
CuCl$_4$ or CuCl$_2$O$_2$ plaquettes are shown in
Fig.~\ref{orbitals}. These plaquettes, relevant for the magnetic
couplings, form edge-shared chains in CuCl$_2$, whereas they are
isolated in the hydrated system.

Naturally, this ``orbital switching'' induced by the crystal water
implies a change of the coupling regime from 1D to 3D. To study
these changes on a quantitative level, we constructed a TB model 
(Fig.~\ref{dos}c) for both compounds and fitted it to the
relevant LDA bands (Fig.~\ref{dos}e,f). The  leading
transfer integrals $t_i$ are given in Table~\ref{table}. The
calculated $t_i$ confirm the intuitive picture that corresponds to
Fig.~\ref{orbitals}: CuCl$_2$ shows quasi 1D dispersion along the
chain with dominating NNN hopping $t_2$, and a considerably weaker
coupling between the chains, while the coupling between adjacent
layers is very small. On the other hand, the changed plaquette
arrangement in CuCl$_2$$\cdot$2H$_2$O, induced by the crystal water, leads 
to a strongly reduced band width (Fig.~\ref{dos}d) and correspondingly 
small isotropic (3D) transfer integrals (Table~\ref{table}).

For the strongly correlated limit ($U_{\mathrm{eff}} \gg t_i$) at half filling, the TB
model can be mapped via a Hubbard model to a Heisenberg model with
resulting AFM exchange couplings $J_i=4t_i^2/U_{\mathrm{eff}}$ where $U_{\mathrm{eff}}$
is the correlation in the effective one-band description. Depending on
the choice of $U_{\mathrm{eff}}$ within a reasonable range \cite{rosner01, rosner97, drechsler} ($U_{\mathrm{eff}} =
3.5 - 4$ \,eV) this leads to exchange constants $J_i = 3 - 5$\,K for the
three leading couplings in CuCl$_2$$\cdot$2H$_2$O (Table~\ref{table}). 
These $J$'s are perfectly in line with the
experimentally observed $T_N = 4.3$\,K as could be expected for an
almost isotropic 3D coupling. For CuCl$_2$, the leading NNN $t_2$
results in $J_2 = 160 - 180$\,K. Although of the correct order of
magnitude compared to $T_{\mathrm{max}}$ and $\Theta_{\mathrm{CW}}$, this would clearly
exceed the ``overall AFM coupling'' in the compound without additional
FM interactions. Sizable FM interactions are typical for close to
90$^\circ$ bond angles according to the Goodenough-Kanamori-Anderson rules \cite{GKA} and
therefore expected for the NN $J_1$ in CuCl$_2$.  For a quantitative
estimate of the FM contributions to the leading $J$'s we apply LSDA+$U$
calculations for different spin arrangements in magnetic super cells.

Mapping the resulting total energy differences to the
Heisenberg model, we obtain the following total exchange integrals
($U_d = 7 \pm 0.5$\,eV) \footnote{The chosen region for $U_d$ covers the
experimental NN exchange $J_1$ in La$_2$CuO$_4$.}: $J_1 =
-(150 \pm 10)$\,K, $J_2 = 155 \pm 25$\,K and $J^{\mathrm{ic}}_1 =
35 \pm 5$\,K. Within the error bars, the latter two agree very well
with the $J$ values calculated from the corresponding $t$'s of the TB
approach,\footnote{This good agreement justifies {\it a posteriori}
the transfer of the $U_{\mathrm{eff}}$ and $U_d$ values from the CuO$_2$ chain
compounds.} indicating that FM contributions beyond NN are rather
small. In contrast, we find a large FM contribution of about 175\,K to
$J_1$ as expected from the Cu-Cl-Cu bond angle of 93.6$^\circ$. The
size of the FM contribution to $J_1$ fits well to related
edge-shared CuO$_2$ chain compounds.\cite{drechsler, nitzsche}

\begin{table}
\begin{ruledtabular}
\begin{tabular}{l|c c c c c }
$t_i$/meV &$t_1$&$t_2$ &$t^{\mathrm{ic}}_1$ &$t^{\mathrm{ic}}_2$ &$t^{\mathrm{il}}_1$ \\
 \hline
CuCl$_2$ & 34 & 117& 61 & -19 & 8 \\
 CuCl$_2$$\cdot$2H$_2$O & 17 & 6 & 4 & 16 & 20 \\
\end{tabular}
\end{ruledtabular}
\caption{\label{table} Calculated leading hopping integrals $t_i$ for the effective TB model shown in Fig. \ref{dos}c.}
\end{table}

The resulting leading exchange interactions confirm the intuitive
picture of a quasi 1D chain model compound with small inter-chain
coupling, very similar to
LiCu$_2$O$_2$.\cite{gippius,masuda,drechsler05} Therefore, the
magnetic ground state is mainly determined by the ratio
$\alpha {\equiv} J_2/J_1$ of the frustrating main interactions along the
chains. For CuCl$_2$, we find $\alpha = -(1.0 \pm 0.1)$ and predict a
ground state well in the helical ordered region of the
$J_1$--$J_2$-phase diagram.

Thus, the dehydration of CuCl$_2$$\cdot$2H$_2$O to CuCl$_2$ leads to a
drastic change of the coupling regime from 3D to quasi 1D and a
completely different ground state that can be traced back to a switch
of the magnetically active orbital.

\subsection{Model analysis of magnetic susceptibility}
For an independent evaluation of the leading exchange interactions in
CuCl$_2$ we simulated $\chi(T)$ within a spin-1/2 $J_1$--$J_2$
Heisenberg model for various $J_2/J_1$ ratios using the TMRG technique
and fitted the resulting $\chi^{*}(T/J)$ curves to the measured
$\chi(T)$. Rather typical for the $J_1$--$J_2$ Heisenberg
model,\cite{masuda,drechsler05} we find two possible solutions for the
fit: (i) $\alpha = +3.0$ with $J_1 = 120$~K and $J_2 = 40$~K (AFM solution)
and (ii) $\alpha = -1.5$ with $J_1 = -90$~K and $J_2 = 135$~K (FM
solution). The FM solution (ii) is in rather good agreement with the
estimates from our {\it ab-initio} calculations. The corresponding
fits are shown in Fig.~\ref{sus}. The AFM solution (i) can be
discarded regarding our calculational results and the close to
90$^\circ$ Cu-O-Cu bond angle.

In a naive approach, using the relation $\Theta_{\mathrm{CW}}\approx
1/2(J_1+J_2+J_{\mathrm{ic}})$,\cite{johnston00} the theoretically estimated
$\Theta_{\mathrm{CW}}^{\mathrm{theo}}\approx$~+30 $\pm 10$\,K 
seems to be inconsistent with the experimental
$\Theta_{\mathrm{CW}}\approx +100$\,K.  Thus, we choose a more
sophisticated procedure performing ED studies for the FM
solution,\footnote{Even for medium size clusters ($N = 16$), ED
  perfectly describes the high-temperature region of $\chi(T)$ which
  obeys a Curie-Weiss law.} which yield the expected
$\Theta_{\mathrm{CW}}\approx1/2(J_1+J_2)$, but only at high
temperatures $T> 10 J_2$.  Since this high temperature region is
inaccessible to experiments, we choose a temperature window of the
Curie-Weiss fit in order to obtain $\Theta_{\mathrm{CW}}$ for the
highest temperatures measured ($225$\,K $< T < 300$\,K corresponding
to $1.7J_2 < T < 2.2J_2$). Using $J_1$ and $J_2$ from the TMRG, we
obtain $\Theta_{\mathrm{CW}}^{\mathrm{theo}} = +72$\,K.  The remaining
difference between $\Theta_{\mathrm{CW}}^{\mathrm{theo}}$ and
$\Theta_{\mathrm{CW}}^{\mathrm{exp}}$ of about 30\,K corresponds in
very good agreement to the inter-chain coupling $J_{\mathrm{ic}}$ of
about 35\,K neglected in the ED simulations.

\subsection{Thermochemical properties}
The dramatic effect of the dehydration of CuCl$_2$$\cdot$2H$_2$O
certainly raises the question whether this is really a typical
dehydration process or if the involvement of the crystal water related
oxygen in the magnetic exchange via covalent Cu-O bonds is an
indication for a chemical reaction on a different energy scale.  The
dehydration enthalpy $\Delta H^{0}_{\mathrm{dehyd.}} = 117$\,kJ/mol of
CuCl$_2$$\cdot$2H$_2$O is a typical (small) value for this class of
materials.\cite{taniguchi, lumpkin} Our {\it ab-initio} estimate for
the dehydration enthalpy yields $\Delta H^{0}_{\mathrm{dehyd.}} =
95$\,kJ/mol which is in quite good agreement with the measured value,
providing additional confidence to the reliability of the
calculational procedure.

\section{Summary}
In summary, our joint theoretical and experimental study provides a
consistent explanation for the fundamentally different magnetic
properties of CuCl$_2$$\cdot$2H$_2$O and CuCl$_2$. Whereas
CuCl$_2$$\cdot$2H$_2$O is a quite isotropic 3D AFM with small exchange
couplings due to the orientation of neighboring isolated CuCl$_2$O$_2$
plaquettes, CuCl$_2$ can be well understood in terms of a 1D FM-AFM
$J_1$--$J_2$ chain model.  This extension of the originally,
critically discussed 1D AFM-NN only model \cite{stout} re-establishes
the pronounced 1D nature of the magnetism in CuCl$_2$.  The dramatic
change of magnetic properties between both compounds can be traced
back to a switch of the magnetically active orbital induced by crystal
water. From our {\it ab-initio} calculations and model studies of the
measured susceptibility using the transfer-matrix density-matrix
renormalization-group and exact diagonalization techniques we predict
a helical ground state and likely related multiferroic behavior for
CuCl$_2$ driven by strong in-chain frustration originating from FM
nearest neighbor and AFM next-nearest neighbor exchange interactions.
Our study reveals that crystal water can have crucial influence on the
electronic and magnetic properties of low dimensional magnets. More
general, our work emphasis that a transfer of model parameters from
seemingly closely related systems is rather dangerous. The unaware
neglect of this fact will lead to an at best inaccurate description 
of the physical properties in many cases.

We acknowledge R. Kremer for valuable discussions, T.~Xiang for
providing the TMRG code, and S. M\"uller for supporting DTA and DSC studies.\\
{\it Note added in revision: While revising our manuscript, we learned that our
magnetic model and predicted ground state was confirmed by an independent study.\cite{kremer}}

\end{document}